\documentclass[aps,prd,twocolumn,nofootinbib,floatfix]{revtex4-1}
\usepackage{setspace}
\usepackage{amsthm}
\theoremstyle{definition}

\usepackage{graphicx}
\usepackage{dcolumn}
\usepackage{multirow}
\usepackage{subfigure}
\usepackage{times,mathptm}
\usepackage{float}
\usepackage{color}
\usepackage{amsmath,amsfonts}
\usepackage{mathptmx}
\usepackage{mathrsfs}
\usepackage{bbm}
\usepackage{bm}
\usepackage{xfrac}

\newcommand{\beq}{\begin{equation}}
\newcommand{\eeq}{\end{equation}} 
\newcommand{\bea}{\begin{eqnarray}}
\newcommand{\eea}{\end{eqnarray}}

\newcommand{\pbar}{\overline{\psi}}
\renewcommand{\d}{\delta}

\renewcommand{\b}{\beta}

\newcommand{\tr}{\text{Tr}}

\newcommand{\vx}{{\vec{x}}}
\newcommand{\vy}{{\vec{y}}}

\newcommand{\n}{\nu}
\newcommand{\m}{\mu}

\newcommand{\g}{\gamma}

\newcommand{\s}{\sigma}

\newcommand{\dg}{\dagger}
\newcommand{\non}{\nonumber}

\newcommand{\rf}[1]{(\ref{#1})}
\newcommand{\ra}{\rightarrow}

\renewcommand{\vec}[1]{\bm #1}
\usepackage{ulem}

\begin{document}

\title{Unphysical properties of the static quark-antiquark four-point correlator in Landau gauge} 

\bigskip
\bigskip

\author{Jeff Greensite and Evan Owen}
%\singlespacing
\affiliation{Physics and Astronomy Department, San Francisco State
University,   \\ San Francisco, CA~94132, USA}
\bigskip
\date{\today}
\vspace{60pt}
\begin{abstract}

\singlespacing
 
   We consider the four point connected correlator representing a static quark-antiquark pair separated
by a spatial distance $R$, propagating for a Euclidean time $T$. This function is computed by lattice Monte Carlo
in SU(2) pure gauge theory at lattice couplings $\b=2.2$ and $\b=2.5$ in both Coulomb and Landau gauges.   The Coulomb gauge
correlator is well behaved, and is dominated at large $T$ by a state whose energy grows linearly as $ \s R$, with $\s$ the known asymptotic string tension.  The connected correlator in Landau gauge behaves differently.  At intermediate $R$ there is
clear evidence of a linear potential, but the corresponding string tension extrapolates to zero at large $T$.  At large $R$ the connected correlator becomes negative; moreover there are strong finite size effects.   These numerical results suggest that unphysical states dominate the large Euclidean time behavior of this Landau gauge correlator. 
\end{abstract}

\maketitle

\singlespacing
%\begin{widetext}
\section{\label{intro}Introduction}

     It is well known that the Landau gauge gluon propagator, as computed in lattice Monte Carlo simulations, violates
reflection positivity  \cite{Bowman:2007du,Cucchieri:2004mf} and this fact is viewed by some as indicative of gluon confinement.  Then it is of interest to
ask whether anything similar happens in Landau gauge quark-antiquark connected four point functions.  One expects that there
are poles in the connected four point functions corresponding to single meson states, and these poles should have a positive residue.  This is, in fact, the starting point of the Bethe-Salpeter approach.
But  equal-times quark-antiquark operators, at distinct spatial points, do not create BRST invariant states, and in
any case both BRST invariance and reflection positivity are problematic in Landau gauge at the non-perturbative level, as we will discuss further below.   So there is at least a possibility that Landau gauge quark antiquark four point
functions exhibit unphysical behavior at large spacetime separations.

    To investigate this possibility, we simplify matters as much as possible.  We consider only the four point functions corresponding to static quarks and antiquarks with spatial separation $R$, evolving for a Euclidean time $T$, evaluated in pure SU(2) gauge theory.  Since the quarks are static, this boils down to evaluating the connected correlator of Wilson lines
\bea
           G(R,T) &=&   {1\over 2} \left\langle L_T^{ab}(\vx) L_T^{\dg ba}(\vy)] \right\rangle 
                               -   {1\over 2}  \left\langle L_T^{ab}(\vx) \right\rangle \left\langle L_T^{\dg ba}(\vy)] \right\rangle  \non \\
                        &=& \left\langle  {1\over 2} \tr [L_T(\vx) L_T^\dg(\vy)] \right\rangle 
                        -  \left\langle  {1\over 2} \tr L_T(\vx)\right\rangle^2 \ ,
\eea
where $L_T(\vx)$ is a timelike Wilson line on the lattice of length $T$, i.e.
\beq
            L_T(\vx) = U_4(\vx,1) U_4(\vx,2)... U_4(\vx,T) \ ,
\eeq
and we have used the fact that, as a consequence of the remnant symmetry under spacetime independent gauge transformations $g(\vx,t)=g$ which exists in Landau gauge
\beq
             \left\langle L_T^{ab}(\vx) \right\rangle = {1\over 2} \left\langle \tr L_T \right\rangle \d^{ab} \ .
\eeq 
We note that in Coulomb gauge there is a remnant symmetry under time-dependent gauge transformations
$g(\vx,t)=g(t)$, and as a result
\beq
           \left\langle L_T^{ab}(\vx) \right\rangle = 0 \ .
\eeq

    In the large Euclidean time limit, $G(R,T)$ in Coulomb gauge should be dominated by the lowest energy
eigenstate of the Coulomb gauge Hamiltonian $H_{Coul}$ containing a static quark-antiquark pair, while $G(R,T)$ in Landau
gauge should likewise be dominated by the lowest energy eigenstate of the BRST Hamiltonian $H_{BRST}$.  But are 
these the same states?  That is the question which we will try to address here numerically.

\section{Results}

    Gauge-fixing is accomplished by the standard over-relaxation method, which applies, in each gauge-fixing sweep, an (over-relaxed) gauge transformation at each site, aiming
to maximize the quantity
\beq
            R = \sum_{\vx} \sum_{i=1}^d \tr[U_i(x)] \ ,
\eeq
where $d=3$ and $d=4$ for Coulomb gauge and Landau gauge respectively.  After each gauge-fixing sweep we calculate the fractional reduction in $R$ compared to the previous sweep.  The gauge-fixing loop ends when the fractional reduction in $R$ falls below $10^{-10}$.

\begin{figure}[t!]
 \includegraphics[scale=0.7]{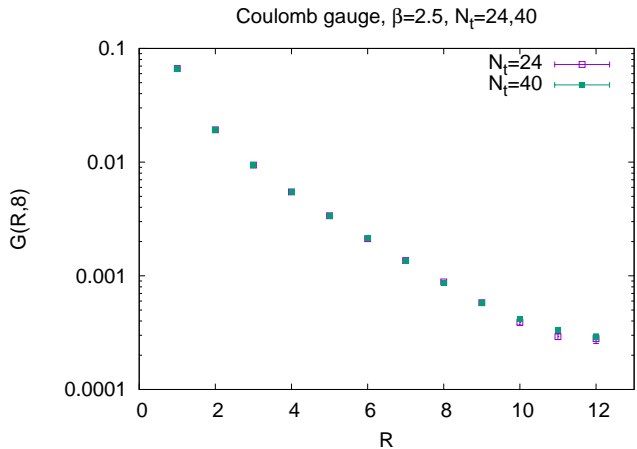}
\caption{Comparison of  a logarithic plot of $G(R,8)$ vs.\ $R$, for lattice volumes $24^3\times N_t$ and $N_t=24$ and $N_t=40$ lattice spacings. There is little difference in the two sets of data, as expected.} 
\label{compare}
\end{figure} 

\begin{figure}[t!]
 \includegraphics[scale=0.7]{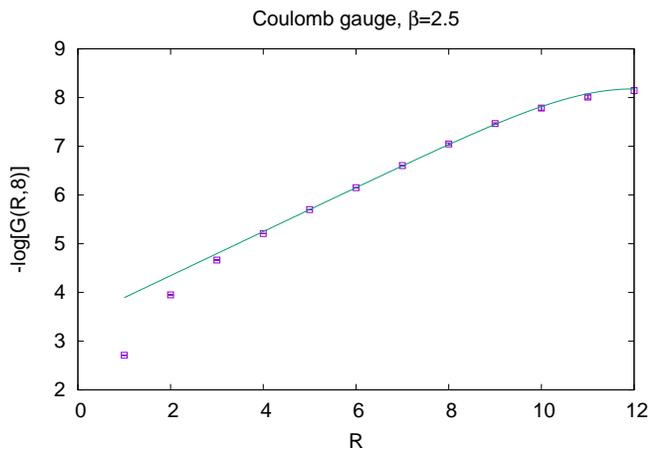}
\caption{Best fit of -log[eq.\ (7)], with $N_s=24$, to the data for $-$log$[G(R,8)]$.} 
\label{T8fit}
\end{figure} 

\begin{figure}[h!]
 \includegraphics[scale=0.7]{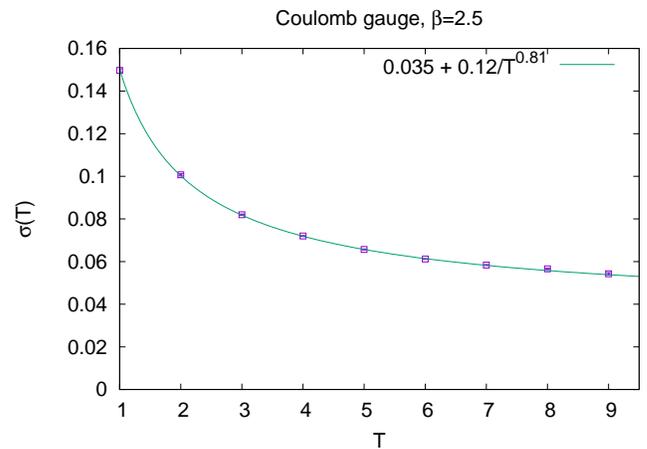}
\caption{String tension $\s(T)$ vs.\ $T$, together with a best fit.  The data extrapolates
to the known asymptotic string tension of $\s=0.035$ in lattice units.} 
\label{sigmaC}
\end{figure} 

\subsection{Coulomb gauge}

   We begin with results for $G(R,T)$ in Coulomb gauge.  This type of calculation is not really new; the first results of this
kind were obtained in ref.\ \cite{Marinari:1992kh}, and later in \cite{Greensite:2003xf} and \cite{Heinzl:2007cp}.  
We have included them here in order to make a comparison with the Landau gauge results to follow.

    At large $R,T$ the $G(R,T)$ correlator is expected to be well approximated by a sum of terms falling exponentially
with both $R$ and $T$.  As an ansatz to extract an ``effective'' string tension $\s(T)$ at fixed $T$, we consider fitting the
large $R$ data to a single exponential
\beq
          G(R,T) \approx c(T) e^{-\s(T) R T} \ .
\eeq
Assuming this gives a good fit to $G(R,T)$, we can then extrapolate $\s(T)$ to $T\ra \infty$, where it is expected to converge to the usual asymptotic string tension. 

However, on the lattice we must allow for periodic boundary conditions.  If the quark-antiquark separation is parallel to the $x, y$ or $z$ axes, and the lattice is $N_s$ spacings in any of the space directions, then it is better to fit $G(R,T)$ to
\beq
         G(R,T) \sim c(T) \Bigl( e^{-\s(T) R T} + e^{-\s(T) (N_s-R) T} \Bigr)  \ .
\label{pbc}
\eeq
By fitting the logarithm of the data for $G(R,T)$ vs.\ $R$,  at large $R$ and fixed $T$, to the logarithm of the right-hand side of \rf{pbc}, we can extract the string tension $\s(T)$.  

    We have carried out these fits at $\b=2.5$ on a $24^3 \times 40$ lattice.  The time asymmetry is actually irrelevant in our Coulomb gauge data, as can be seen by comparing $G(R,T)$ at $T=8$ computed on a $24^3 \times N_t$ lattice with 
$N_t=24$ and $N_t=40$.  The comparison is shown in Fig.\ \ref{compare}, at it is clear that the difference due to increasing
$N_t$ from 24 to 40 is essentially negligible, as one would expect.  We have nevertheless carried out simulations at
$N_t=40$ in order to compare the Coulomb gauge data with the Landau gauge data on the same lattice volume.

   The fit of our data to eq.\ \rf{pbc} is illustrated for $G(R,T)$, again at $T=8$, in Fig.\ \ref{T8fit}, where the fitting region was the range $R > 4$.  The figure is representative of similar fits from $T=1$ to $T=9$.  We plot the values of $\s(T)$ extracted
from these fits in Fig.\ \ref{sigmaC}.  The data is found to closely follow the curve
\beq
         \s(T) = \s_{\infty} + {0.12 \over T^{0.81} }  ~~~ \mbox{where} ~~~  \s_\infty =  0.035(1) \ .
\eeq
The asymptotic value $\s_\infty$ agrees within errorbars with the SU(2) string tension at $\b=2.5$ reported in \cite{Bali:1994de}.
    
    So far there are no surprises.  These results are consistent with expectations.

\subsection{Landau gauge}

\begin{figure}[t!]
\subfigure[~]  % caption for subfigure a
{   
 \label{b22A}
 \includegraphics[scale=0.65]{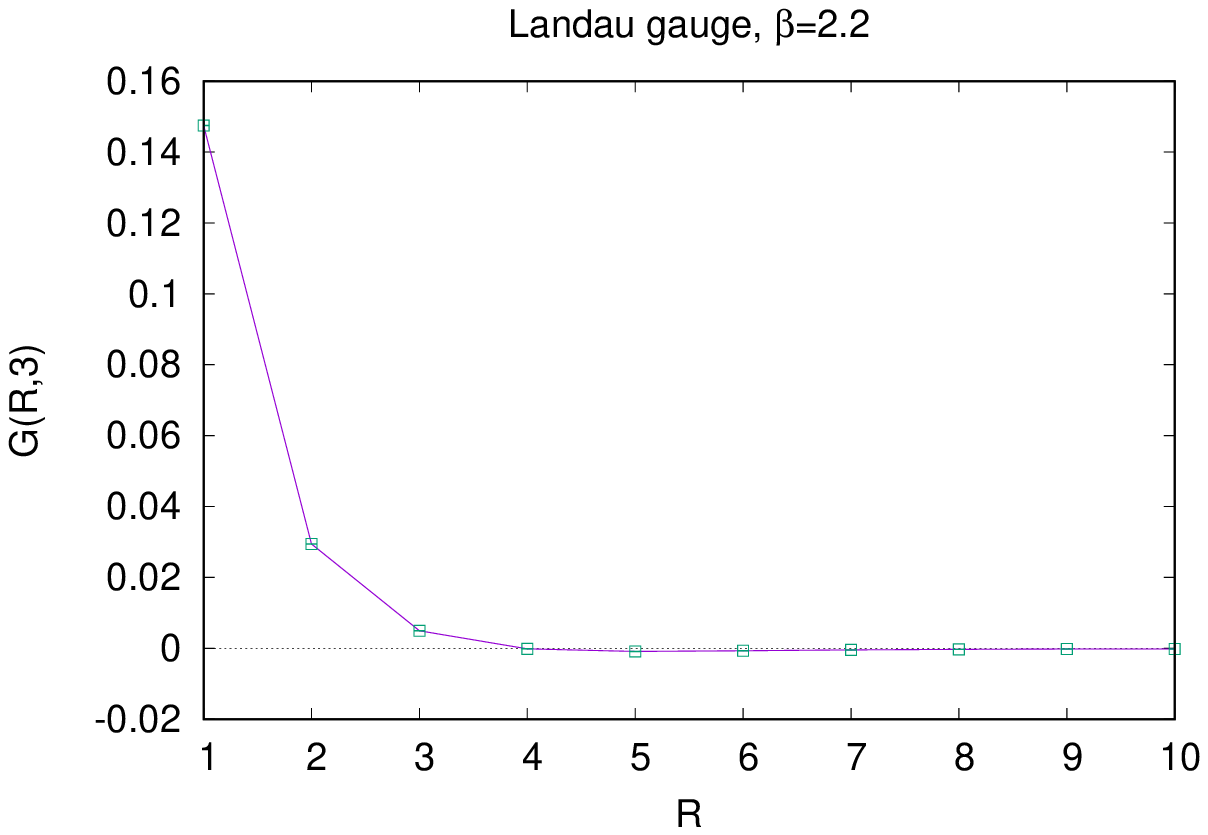}
}
\subfigure[~]  % caption for subfigure a
{   
 \label{b22B}
 \includegraphics[scale=0.65]{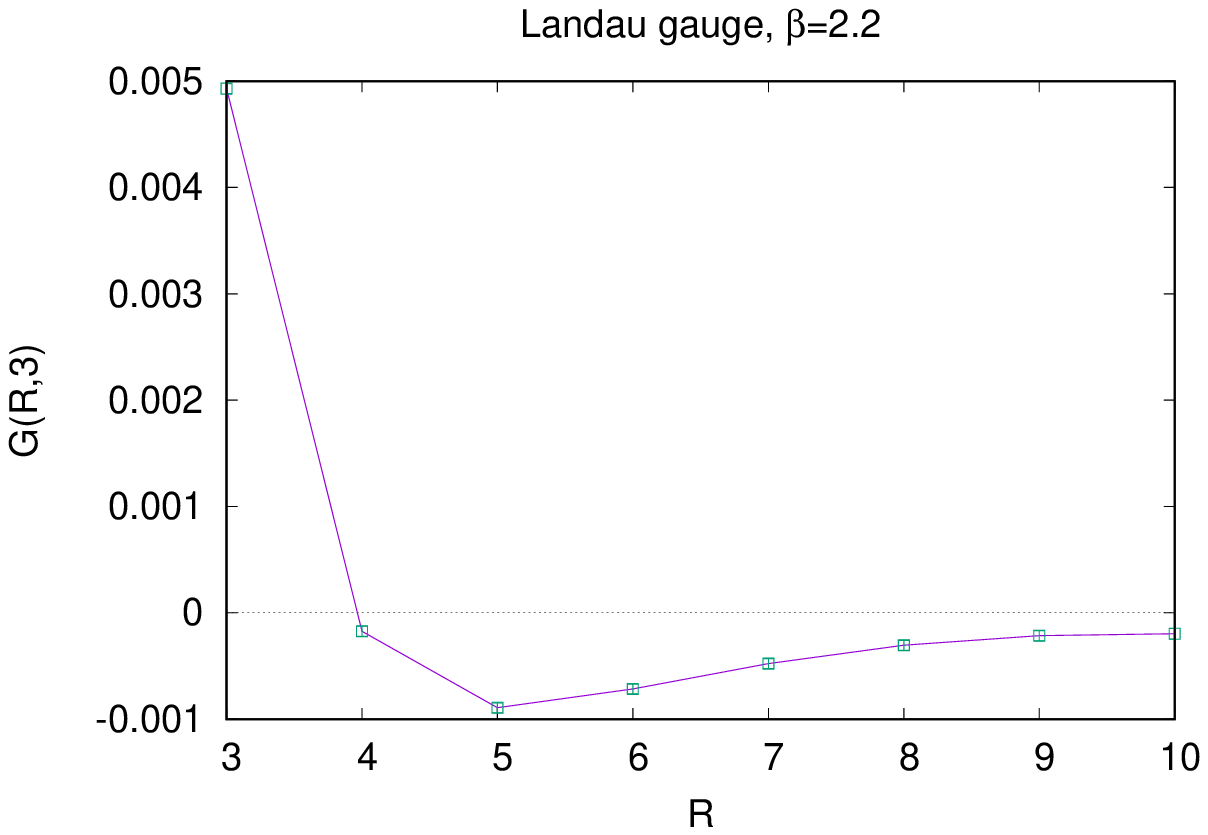}
}
\caption{Landau gauge connected correlator $G(R,T)$ vs.\ $R$ at $\b=2.2$ and $T=3$ on a $20^4$ lattice. (a) $G(T,4)$ in the
full range of $R$.  (b) closeup in the range $R>2$. Note the violation of positivity in this range.} 
\label{b22}
\end{figure}

\begin{figure}[t!]
\subfigure[~]  % caption for subfigure a
{   
 \includegraphics[scale=0.65]{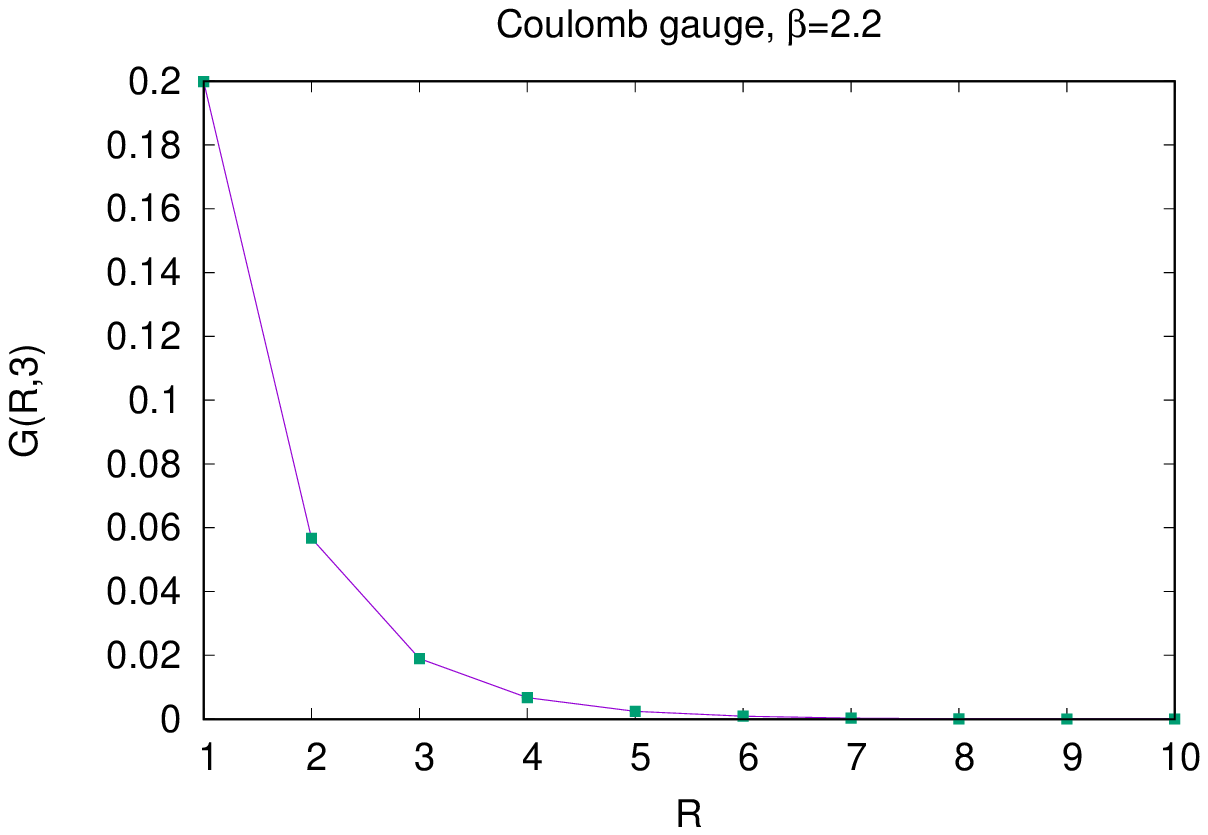}
}
\subfigure[~]  % caption for subfigure a
{   
 \includegraphics[scale=0.65]{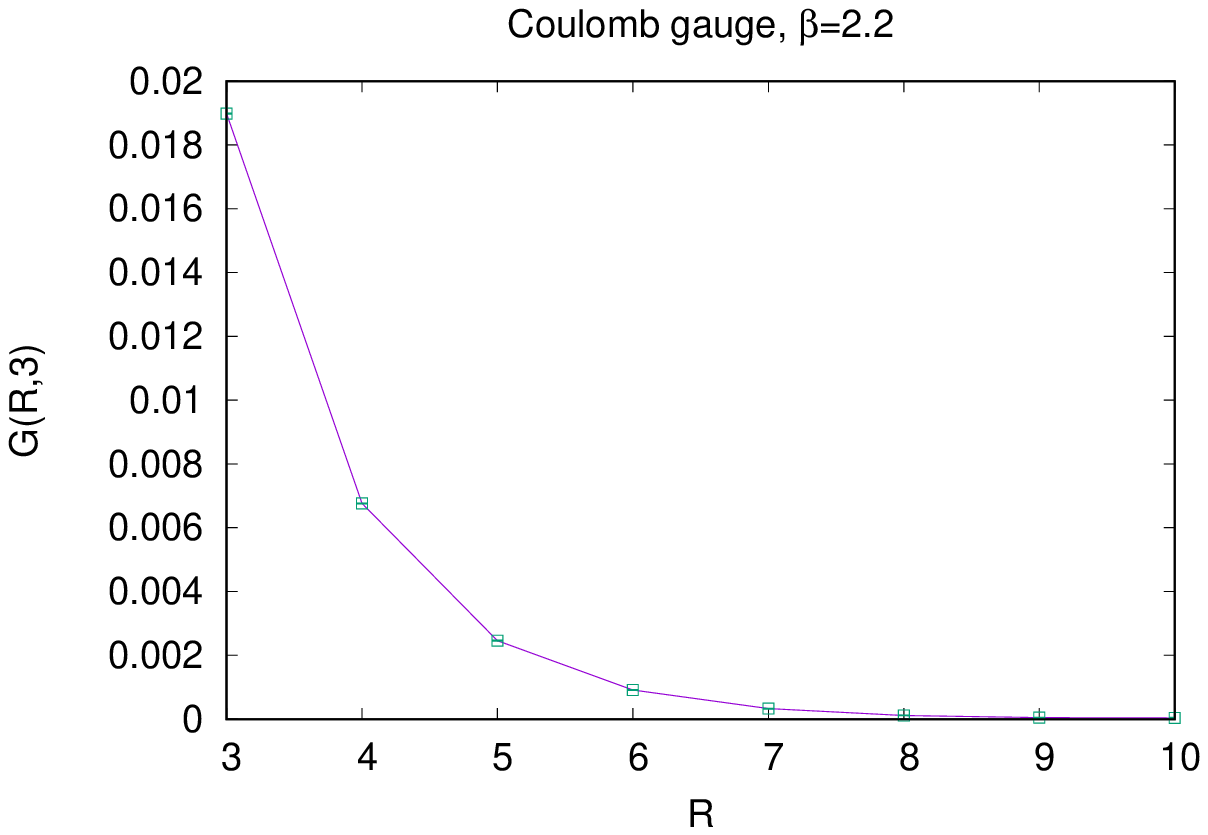}
}
\caption{Same as Fig.\ \ref{b22}, but in Coulomb gauge.} 
\label{b22C}
\end{figure}

   In Fig.\  \ref{b22A} we display $G(R,T)$ vs.\ $R$ in Landau gauge at $T=3$ and an intermediate coupling strength of $\b=2.2$.  The lattice volume is $20^4$.  A closeup of the data at separations $R \ge 3$ is shown in Fig.\ \ref{b22B}.  From this figure it
is clear that the correlator violates positivity from $R=4$ onwards.  We find a similar positivity violation in all plots of $G(R,T)$
vs.\ $R$ at all $T$.  For comparison we show in Fig.\ \ref{b22C} the same plots at the same $\b=2.2$ and lattice volume in Coulomb gauge

\begin{figure}[t!]
 \includegraphics[scale=0.7]{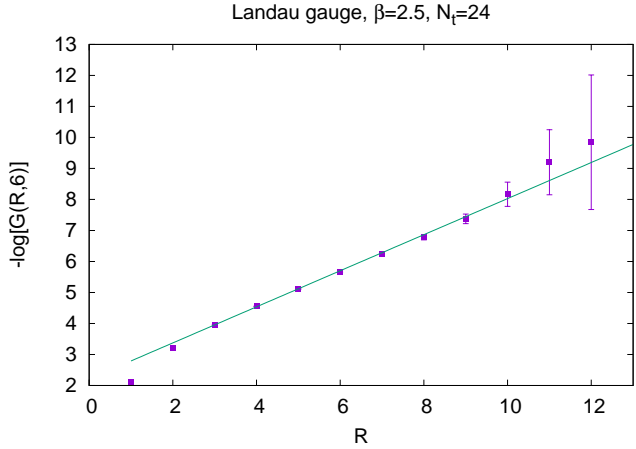}
\caption{Data for $-$log$G(R,6)$ vs.\ R in Landau gauge at $\b=2.4$ on a $24^4$ lattice, together with
a linear fit through the larger $R$ values.} 
\label{t24lin}
\end{figure}

\begin{figure}[htb]
 \includegraphics[scale=0.7]{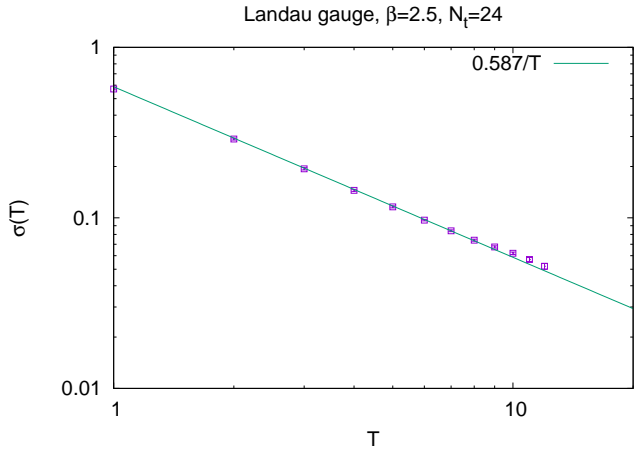}
\caption{String tension in Landau gauge, in lattice units, at $\b=2.5$, obtained from fits to
$G(R,T)$ data obtained on a $24^4$ lattice volume. } 
\label{tension24}
\end{figure}

   At first sight this positivity violation seems to disappear at $\b=2.5$.  In Fig.\  \ref{t24lin} we display $G(R,T)$ at $T=6$,
shown here on a logarithmic scale.  It is clear that for $3 \le R \le 9$ the data on a log plot is fit pretty well
by a straight line, and this holds true for all $T$ up to $T=12$.  
Therefore, at $\b=2.5$ on a $24^4$ lattice, we can follow the
previous procedure in Coulomb gauge, and extract a $T$ dependent string tension $\s(T)$ from a fit of the data to
\beq
            G(R,T) \approx e^{-\s(T) R T}  ~~~  R \ge 3 \ .
\eeq
Fig.\ \ref{tension24} is a plot of $\s(T)$ vs.\ $T$ on a log-log plot.  Unlike Coulomb gauge, the data is fit fairly well by
\beq
            \s(T) \approx {0.587 \over T} \ ,
\eeq
which means that $\s(T)$ extrapolates to zero as $T\ra \infty$.   The implication is that the Green's function is dominated,
at large Euclidean times, by a state with zero string tension, i.e.\ an unphysical state.

\begin{figure}[htb]
\subfigure[~]  % caption for subfigure a
{   
 \label{b25A}
 \includegraphics[scale=0.65]{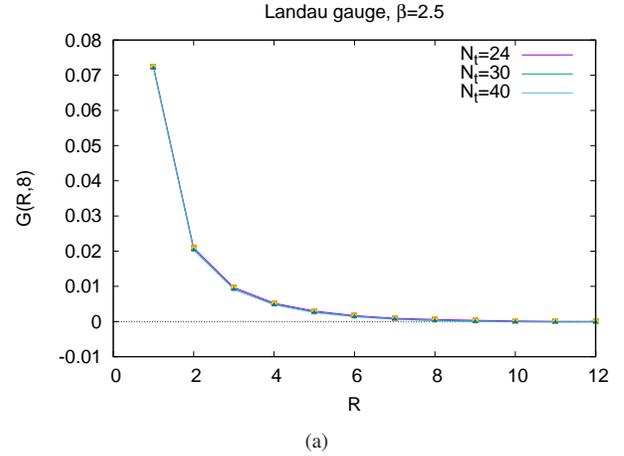}
}
\subfigure[~]  % caption for subfigure a
{   
 \label{b25B}
 \includegraphics[scale=0.65]{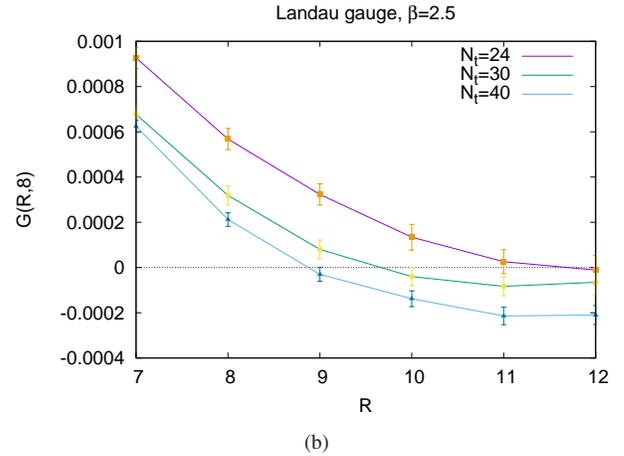}
}
\caption{Landau gauge connected correlator $G(R,T)$ vs.\ $R$ at $\b=2.5$ and $T=8$ on a $24^3 \times N_t$ lattice,
with $N_t=24,30,40$. (a) $G(T,8)$ in the
full range of $R$.  (b) closeup in the range $R>6$, where we observe positivity violation at $N_t=30,40$.}  
\label{b25}
\end{figure}

     We notice, however, that in  Fig.\  \ref{t24lin} and in all other plots of $G(R,T)$ vs.\ $R$ at constant $T$, there is no
evidence of the ``flattening out''  of the data at the largest two or three values of $R$,
which would have been expected due to
periodic boundary conditions.  In fact, and in contrast to Coulomb gauge, the data points at $R=10, 11$ seem to even lie
{\it above} the straight line fit, albeit there are large error bars.  To investigate this further, we have increased the length of the lattice in the time direction to $N_t=30$  and  $N_t=40$, while keeping the extension in the space directions fixed at 24 lattice spacings.  When we do that, we find that the positivity violation found at $\b=2.2$ reappears in $G(R,T)$ {\it at all $T$} for $R>8$, as shown, e.g., in Fig.\ \ref{b25} at $T=8$.  Evidently, apart from positivity violation, the Landau gauge correlator is subject to severe finite size effects.

\begin{figure}[t!]
 \includegraphics[scale=0.7]{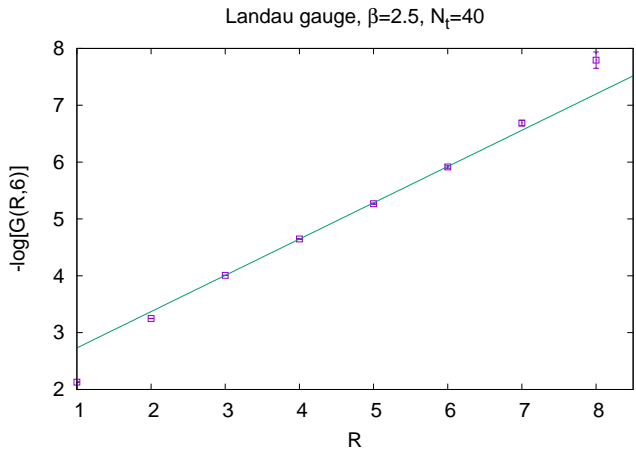}
\caption{Data for $-$log$G(R,6)$ vs.\ R in Landau gauge at $\b=2.4$ on a $24^3 \times 40$ lattice, together with
a linear fit in the range $3\le R \le 7$.   Note that $G(R,T)<0$ for $R>8$, so those points cannot be displayed in this figure.} 
\label{t40lin}
\end{figure}

\begin{figure}[t!]
 \includegraphics[scale=0.7]{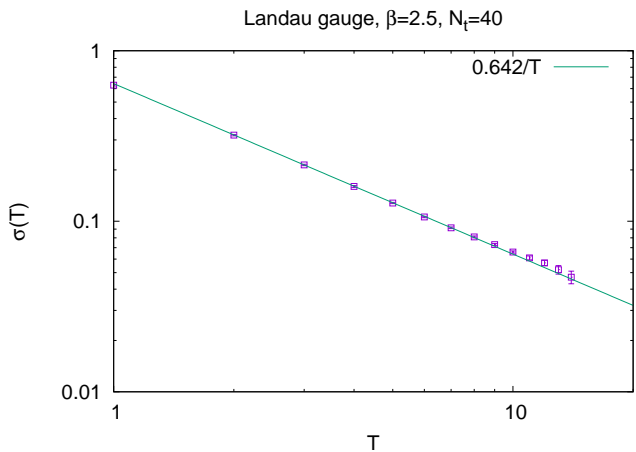}
\caption{String tension in Landau gauge, in lattice units, at ${\b=2.5}$, obtained from fits to
$G(R,T)$ data obtained on a $24^3 \times 40$ lattice volume.   In this case the string tension is extracted from
fits to a restricted range $3\le R \le 7$, due to the non-positivity of the correlator at large $R$.} 
\label{tension40}
\end{figure}

    In an intermediate range of $3 \le R \le 7$, we still observe on the $24^3 \times 40$ lattice a linear rise in 
$-\log[G(R,T)]$ vs.\ $R$, as seen in Fig.\ \ref{t40lin},  and this allows us to extract the effective string tension $\s(T)$ of the corresponding potential in this range.  That string tension is plotted vs.\ $T$ on a log-log scale in Fig.\ \ref{tension40}, and is fairly well fit by
\beq
           \s(T) = {0.642/T} \ .
\eeq
As on the $24^4$ lattice volume, this string tension formally extrapolates to zero at $T \ra \infty$, which of course is the
wrong answer for the energy of a physical state containing a static quark-antiquark pair.  However, we must note that
for $T>12$ the non-positivity affects data points down to $R=6$, and we do not feel justified in extracting a string tension
from only three data points.

\section{Landau gauge and its discontents}

    Obviously it would be wrong to conclude, from the non-positivity of $G(R,T)$ in Landau gauge, that quark-antiquark bound states are absent in the spectrum.  The problem is more likely due to the fact that the relevant quark-antiquark operator in Landau gauge is not a BRST singlet, and moreover that BRST symmetry is itself problematic at the non-perturbative level.

   If the state created by the massive quark-antiquark creation operator $\pbar^{+a}(\vx,t) \psi^{+a}(\vy,t)$ is not annihilated by the
BRST charge operator in a covariant gauge, then it is not a physical state.  It may have an overlap with physical states, but there may also be non-negligible overlaps with negative norm and other unphysical states.   In addition,
BRST symmetry on the lattice is subject to the 0/0 problem pointed out by long ago by Neuberger \cite{Neuberger:1986xz}.  
Let
\beq
          Z = \int DU D\overline{c} Dc e^{-(S+S_{gf})} \ ,
 \eeq
where $S_{gf}$ is the standard BRST gauge-fixing term in a covariant gauge.  Then, as shown in \cite{Neuberger:1986xz}, it follows that $Z=0$.  This is also true if a BRST invariant operator is inserted in the integrand, hence the expectation value of any such observable
is formally 0/0.  The problem has to do with a summation over Gribov copies in covariant gauges, which contribute to the
functional integral with both positive and negative signs.   On the lattice, the gauge-fixing procedure restricts the evaluation 
to Gribov copies within
the first Gribov horizon; i.e.\ to gauge copies which contribute to the expectation values with only positive sign.  But this
restriction itself breaks BRST invariance, as shown numerically by Cuccieri et al.\  \cite{Cucchieri:2014via}.  

   In the absence of BRST invariance, even the usual assumptions underlying reflection positivity are suspect.  Take, for example, the case of the Landau gauge gluon propagator.  What is actually computed on the lattice is the expectation value
\beq
       D^{ab}_{\m \n}(x-y) = \langle [G_L\circ A]^a_\m(x) [G_L \circ A]^b_\n(y) \rangle \ ,
\eeq
where $G_L$ is a gauge transformation which takes the gauge field into some copy of Landau gauge within the
first Gribov horizon.  But $G_L$ is obviously non-local in time, which violates one of the assumptions underlying the usual 
proof of reflection positivity, and this is perhaps the reason for the observed lack of positivity in the Landau gauge gluon
propagator.   If the gauge copies are not restricted to the Gribov region, then one might argue that
time non-locality could be eliminated at the price of introducing ghost fields, i.e.
\beq
          D^{ab}_{\m \n}(x-y) = {1\over Z} \int DA_\m Dc D\overline{c} ~ A^a_\m(x) A^b_\n(y) \exp[-(S+S_{gf})] \ .
\eeq
But this strategy, as already mentioned, runs right into the Neuberger 0/0 problem.  For Landau gauge, and for
covariant gauges in general, the choice is to either break BRST explicitly, or face the 0/0 problem.

    Neither option is attractive.  In lattice simulations the choice is to break BRST symmetry explicitly, which at
least produces a well-defined answer.  But perhaps it is then not surprising that the resulting four-point  Euclidean Green's functions for massive quark antiquark states are found to exhibit unphysical behavior.

\section{Conclusions}

   We have found that the Coulomb gauge four point function $G(R,T)$ corresponding to creation and destruction of a static quark antiquark pair, separated by a spatial distance $R$, behaves as expected:  the correlator  falls off exponentially with $RT$ as in
eq.\ \rf{pbc}, with an effective string tension $\s(T)$ extrapolating,
as $T\ra \infty$, to the known asymptotic string tension.  In contrast, the corresponding connected two point function
in Landau gauge exhibits two pathologies.  First, while $G(R,T)$  does fall off exponentially with $R$ for an intermediate range of $R$, the string tension $\s(T)$ appears to extrapolate to zero at large $T$, indicative of dominance by an unphysical state.  Secondly, at large $R$, the connected four point function is negative, likewise indicating dominance by negative norm states. 

    The first question is whether these types of unphysical behavior persist in quark-antiquark four point functions for quarks with finite mass, and  this will be the next issue to investigate.  Assuming that unphysical behavior persists at finite mass, which we believe is likely, the next question is: do our results pose a problem for the existing Dyson-Schwinger  (DS) and Functional Renormalization Group (FRG) approaches to Landau gauge-fixed QCD, both of which entail the non-perturbative computation of irreducible n-point functions? 
    
     It is difficult to provide a definite answer at the moment.   Some studies, e.g.\ \cite{Eichmann:2016yit},  which combine the Dyson-Schwinger and Bethe-Salpeter 
equations, have had quite some success in treating the low-lying hadron spectrum.  It may be that these approaches somehow avoid the issue of unphysical states, perhaps by concentrating
on n-point functions in the neighborhood of physical poles.  Is it  then possible, within the DS and FRG schemes,
to also uncover the presence of unphysical states in the four point correlation functions?  Or are such states
necessarily absent in these approaches?  Perhaps the truncations which are inevitable in the DS and FRG schemes lose information about unphysical states? (If so, what else might be lost?)  We don't know the answers to any of these questions, but we believe they may be worth further investigation.
    
\bigskip

\acknowledgments{JG would like to acknowledge discussions with Christian Fischer and Jan Pawlowski.  This work is supported by the U.S.\ Department of Energy under Grant No.\ DE-SC0013682.}   

\bibliography{land}

\end{document}